\documentclass[letterpaper]{article} 
\usepackage{aaai25}
\usepackage{times}  
\usepackage{helvet}  
\usepackage{courier}  
\usepackage[hyphens]{url}  
\usepackage{graphicx} 
\usepackage{natbib}  
\usepackage{caption} 

\usepackage{algorithm}
\usepackage{algorithmic}
\usepackage{color}
\usepackage{booktabs}
\usepackage{amsmath}
\usepackage{xspace} 
\usepackage{soul}
\usepackage{amssymb}
\usepackage{multirow} 

\usepackage{xspace}

\newcommand{\mName}{BRIDGE\xspace}

\usepackage{newfloat}
\usepackage{listings}
\DeclareCaptionStyle{ruled}{labelfont=normalfont,labelsep=colon,strut=off} 
\floatstyle{ruled}
\newfloat{listing}{tb}{lst}{}
\floatname{listing}{Listing}
\nocopyright 

\def\showauthors@on{T}
\title{BRIDGE: Bundle Recommendation via Instruction-Driven Generation}
\author{
    Tuan-Nghia Bui,
    Huy-Son Nguyen,
    Cam-Van Thi Nguyen,
    Hoang-Quynh Le,
    Duc-Trong Le
}
\affiliations{

    University of Engineering and Technology, \\ 
    Vietnam National University, Hanoi, Vietnam \\
    \{21020364, huyson, vanntc, lhquynh, trongld\}@vnu.edu.vn    
}

\usepackage{bibentry}

\begin{document}

\maketitle

\begin{abstract}
Bundle recommendation aims to suggest a set of interconnected items to users. However, diverse interaction types and sparse interaction matrices often pose challenges for previous approaches in accurately predicting user-bundle adoptions. Inspired by the distant supervision strategy and generative paradigm, we propose BRIDGE, a novel framework for bundle recommendation. It consists of two main components namely the correlation-based item clustering and the pseudo bundle generation modules. Inspired by the distant supervision approach, the former is to generate more auxiliary information, e.g., instructive item clusters, for training without using external data. This information is subsequently aggregated with collaborative signals from user historical interactions to create pseudo `ideal' bundles. This capability allows BRIDGE to explore all aspects of bundles, rather than being limited to existing real-world bundles. It effectively bridging the gap between user imagination and predefined bundles, hence improving the bundle recommendation performance. Experimental results validate the superiority of our models over state-of-the-art ranking-based methods across five benchmark datasets.
\end{abstract}

\section{Introduction}

Beyond item recommendation, which focuses on suggesting individual items to users, bundle recommendation captures a more nuanced understanding of user behavior in recommending a set of cohesive items for an exclusive intention. The understanding is often built upon leveraging historical user-item, user-bundle interactions, and bundle-item affiliations to learn user preferences \cite{BundleGT2023, ma2024multicbr}. 
This task has gained significant attention in recent years due to its complexity. 
\begin{figure}[t]
\includegraphics[width=0.46\textwidth]{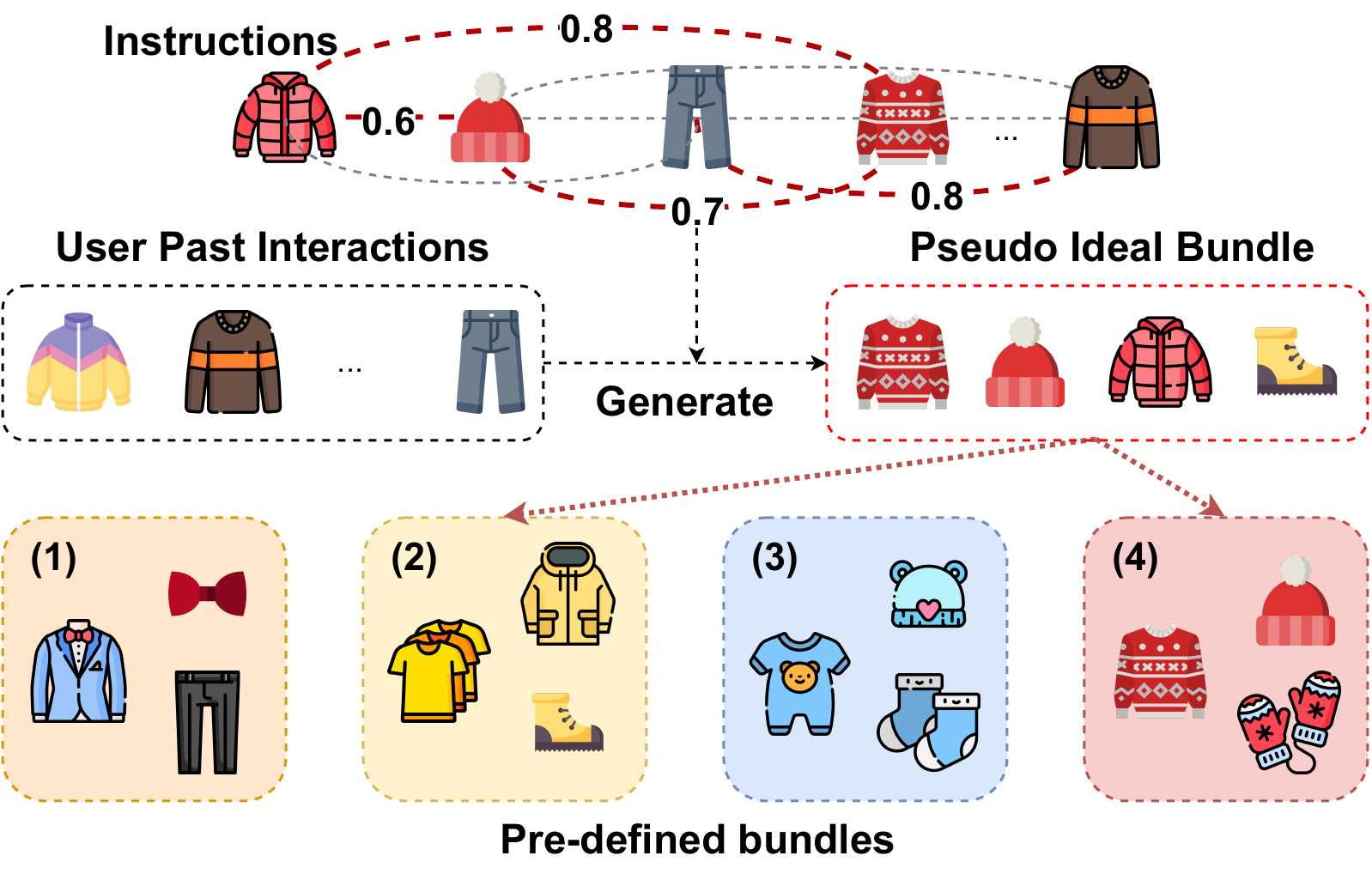}
\caption{An example of instruction-driven bundle recommendation}
\label{insp_fig}
\end{figure}

Exploring user preferences in recommendation systems, especially in the context of bundle recommendation, is a critical and complex challenge. It plays a vital role in improving user experiences across diverse domains \cite{cui2020personalized}.
Most effective graph-based models \cite{chang2020bundle, ma2022crosscbr, zhao2022multi} employ Bayesian Personalized Ranking (BPR) \cite{rendle2012bpr} as the primary objective. These models differentiate between unseen user-bundle interactions (negative samples) and observed user-bundle interactions (positive ones) to rank recommendations by classifying true negatives.    
Regardless of the success of ranking-based methods across various recommendation domains, there exist several serious limitations, which call into question their ability to fully capture and respond to the complexities of user behaviour. As highlighted by \citet{yang2024generate}, one of the primary concerns is the oversimplification of human behaviour in adopting bundles of items. In the other hand, these methods often lack of robustness in the face of data sparsity and noise. In diverse real-world scenarios, user-item interaction data is incomplete, noisy, and sparse, which poses urgent challenges for ranking models in deriving reliable inferences. This issue is even more pronounced in the context of bundle recommendations, where the user-bundle interaction data tends to be even sparser and less consistent. The complexity of bundles, which involve multiple items rather than single ones, adds another layer of difficulty in accurately modeling user preferences. 
When confronted with such limited data, ranking models are prone to over-fitting, relying too heavily on the small amount of available information.

Our research is inspired by the distant supervision strategy, which leverages auxiliary resources to generate silver-standard labeled data for model training \cite{craven1998learning, mintz2009distant}.
However, the effectiveness of distant supervision in bundle recommendation remains largely unexplored due to the scarcity of relevant knowledge bases. Bundle recommendation datasets typically consist of user, bundle, and item IDs, but the lack of information to identify common objects across different datasets makes it more challenging to leverage multiple external data sources to enhance recommendation performance.

To address these challenges, we propose a novel framework named \textbf{\mName{}} for \textbf{B}undle \textbf{R}ecommendation using \textbf{I}nstruction-\textbf{D}riven \textbf{GE}neration, which is inspired by the strategy of distant supervision \cite{craven1998learning, mintz2009distant} and generative retrieval \cite{yang2024generate, rajput2024recommender}. Unlike traditional distant supervision methods that rely on external data sources, \mName{} 
generates silver-standard data through determining clusters of high-correlation items from user-item associations using the \textit{Correlation-based Item Clustering} module. Leveraging historical interacted items of a given user, these clusters are jointly exploited as `instructions' to produce the pseudo `ideal' bundle that aligns with the user preference via the \textit{Pseudo Bundle Generation} module.  The pseudo bundle helps retrieve relevant bundles from predefined options for personalized recommendation within the \textit{Retrieval \& Ranking} module. Figure~\ref{insp_fig} illustrates an intuitive example how \mName{} works. 

To summarize, our main contributions are as follows:
\begin{itemize}
    
    \item We improve the training process for the bundle recommendation task with auxiliary instructions, i.e, correlation-based item clusters, using a distant supervision-based approach without using any external data sources.
    
    \item We propose \mName{}, which uses instructional guidance from historical user interactions and item clusters to generate pseudo 'ideal' bundles to discover relevant candidates from existing bundles for recommendation. To the best of our knowledge, we are the first to present an end-to-end generative approach for bundle recommendation. 
    
    \item We conduct extensive experiments on five publicly-available datasets and achieve significant improvements over all baseline methods on various metrics.
\end{itemize}

The remainder of the paper is organized in the following sections: Related studies for the bundle recommendation task are literally discussed in the 'Related work' section, which is followed by the methodology section to thoroughly described the proposed model \mName{}. We investigate its effectiveness in the 'Experiments' section before summarizing findings in the final section.

\section{Related Work}
\textbf{Bundle Recommendation. }Research on bundle recommendation typically focuses on three main approaches: \textit{factorization methods} decompose interaction matrices into latent factors to predict and enhance bundle recommendations; \textit{graph-based methods} utilize graphs to capture complex relationships between users, items, and bundles, refining recommendations through graph-based techniques; and \textit{generative methods} employ generative models to create pseudo ideal bundles, addressing limitations of traditional ranking approaches by exploring new configurations aligned with user preferences.

\textit{Factorization Methods.} Early bundle recommendation methods, based on the BPR framework \cite{rendle2012bpr}, use user-bundle interactions as positive pairs and sample negative pairs from unobserved interactions. DAM \cite{chen2019matching} is a multi-task framework that recommends both items and bundles using an attention mechanism and shared weights to capture user preferences at both levels.

\textit{Graph-based Recommendation}.
LightGCN~\cite{he2020lightgcn} applies a simple but effective yet graph convolutional operators on the bipartite user-item graph to facilitate item recommendation.
Recently, BGCN~\cite{chang2020bundle} integrates Graph Convolutional Networks with multi-view learning to exploit various interactions.
Addressing the inconsistency between item preference and bundle preference of users, CrossCBR~\cite{ma2022crosscbr}, MultiCBR~\cite{ma2024multicbr} adopt InfoNCE~\cite{gutmann2010noise} to align multi-view user preferences, while MIDGN~\cite{zhao2022multi} models multiple intention hidden in users/bundles.
BundleGT~\cite{BundleGT2023} explores the strategy-aware ability of user/bundle representations. 
CoHeat \cite{jeon2024cold} improve the cold-start problem in bundle recommendation via popularity and curriculum heating.

\textit{Generative Recommendation}.
In traditional item recommendation, PURE \cite{zhou2021pure} uses GANs to generate fake user and item embeddings, covering diverse feature space corners. LARA \cite{sun2020lara} applies multiple generators on item attributes to create pseudo user profiles, with a discriminator classifying real user-item pairs. DreamRec \cite{yang2024generate} employs a diffusion process to reconstruct oracle item embeddings and retrieve recommendations based on similarity to these oracles. TIGER \cite{rajput2024recommender} utilizes Transformer architecture to generate item aspects matching user interests, enhancing item representations with auxiliary data.

\paragraph{\textbf{Distant Supervision.}} Distant supervision is an efficient training strategy applied across various problems and domains, including relation extraction \cite{mintz2009distant, quirk2016distant}, procedural activities recognition \cite{lin2022learning}, and image captioning \cite{qi2024relational}. 
In the context of recommendation systems, distant supervision has been successfully applied for cross-domain recommendation \cite{elkahky2015multi, cao2022contrastive}.

\begin{figure*}[t]
\centering
\includegraphics[width=\textwidth]{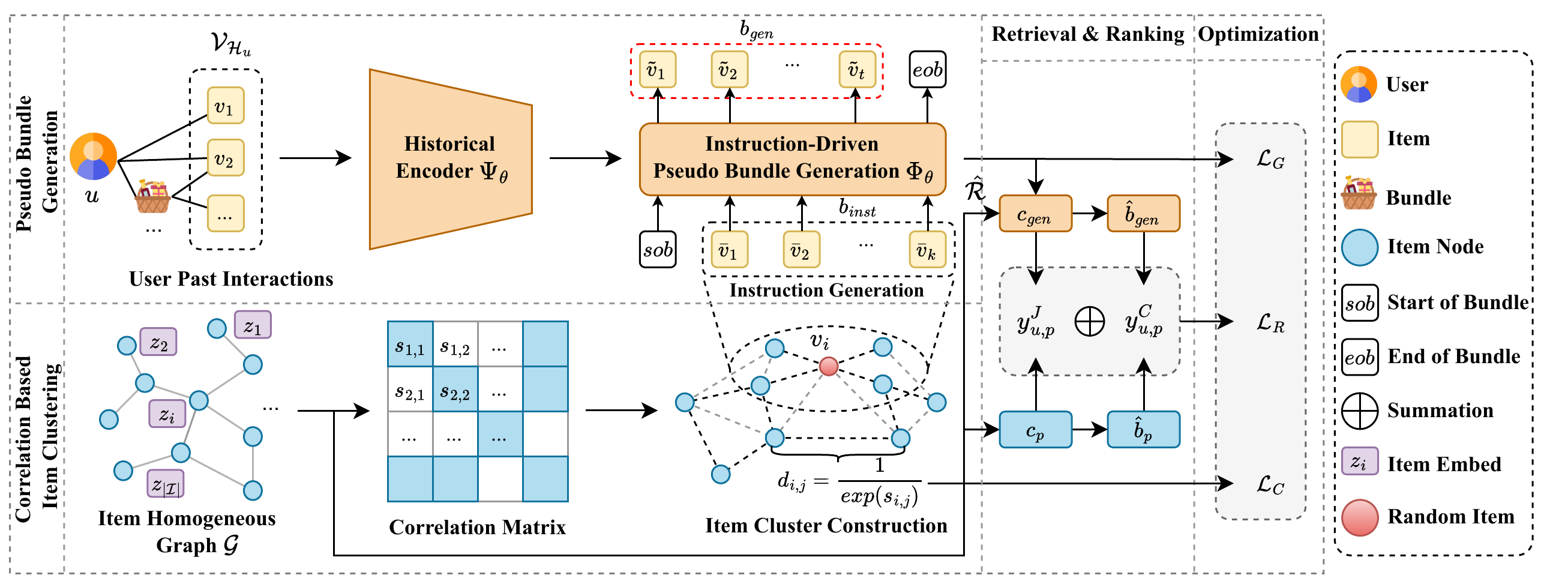}
\caption{The end-to-end architecture of \mName}
\label{overallarchi_fig}
\end{figure*}

\section{\mName{} - Bundle Recommendation via Instruction-Driven Generation}
The overall architecture of \mName{} is shown in Figure~\ref{overallarchi_fig}. It consists of three main components including item-level correlation clustering, pseudo bundle generation and retrieval \& ranking modules. Each component will be thoroughly described in the following subsections. 


\subsubsection{\textbf{Problem Formulation.}} 
Given sets of users $\mathcal{U}=\{u_1, u_2, \dots, u_{|\mathcal{U}|}\}$, bundles $\mathcal{B}= \{b_1, b_2, \dots, b_{|\mathcal{B}|}\}$, and items $\mathcal{V}=\{v_1, v_2, \dots, v_{|\mathcal{V}|}\}$, the observed user-bundle, bundle-item and user-item interactions are respectively represented as three binary matrices $X \in \{0, 1\}^{|\mathcal{U}| \times |\mathcal{B}|}$, $Y \in \{0, 1\}^{|\mathcal{B}| \times |\mathcal{V}|}$ and $Z \in \{0, 1\}^{|\mathcal{U}| \times |\mathcal{V}|}$, where cells with value of $1$ if there exists links between user-bundle, bundle-item or user-item pairs, and 0 otherwise. 
For a given user $u \in \mathcal{U}$ and a predefined bundle $b \in \mathcal{B}$, we seek to compute the  score $y_{u, b}$ that the user $u$ will adopt the bundle $b$ as:
\begin{equation}
    y_{u, b} = g(b, f_\theta(u, X, Y, Z))
\end{equation} where $g$ is a similarity function, and $f_\theta(u, X, Y, Z)$ is the pseudo bundle generation function with a learnable parameter $\theta$, which manifests the preferential representation of $u$. Specifically, our goal is to maximize the similarity between the target bundle and the generated bundle. The top-$K$ bundles with the highest similarity values will be recommended to user $u$.





\subsection{Correlation-based Item Clustering}
We assume that two items interacted by a common set of users via $Z$ may indicate a potential combination within a pseudo bundle, suggesting that they are closely related in the representation space. This pseudo bundle can serve as an instruction signal to guide the bundle generator to create a meaningful bundle.
To verify this hypothesis, we compute the item co-occurrence matrix $C= Z^T \cdot Z, C \in \ \mathbb{R}^{|\mathcal{V}| \times |\mathcal{V}|}$.
Next, we establish an item homogeneous graph $\mathcal{G} = \{\mathcal{V}, E\}$, where $\mathcal{V}$ is the set of nodes and $E = \{e_{i,j}~|~v_i, v_j \in \mathcal{V}\}$ denotes the edge set, $e_{i, j} = 1$ if $C(i, j) > 0$, and 0 otherwise.

In order to learn the item latent embedding of a given item $v_i$ using topological information, we employ a graph convolution network, i.e.,  LightGCN\cite{he2020lightgcn}, on the item homogeneous graph $\mathcal{G}$. Let us denote $r_i^{(l)}$ is the item latent representation of item $v_i$ at the $l-$th layer. It is derived as:
\begin{equation}
\label{lgcn}
    r_i^{(l)} = \sum_{j \in \mathcal{M}_i} \frac{1}{\sqrt{|\mathcal{M}_i|} \sqrt{|\mathcal{M}_j|}} r_j^{(l-1)},
\end{equation}
where $r_i^{(0)}\in \mathbb{R}^d$ is randomly initialized, ${\mathcal{M}}_i$ represent the neighbor set of item $i$ in $\mathcal{G}$.
The final item latent representation $\hat{r}_i$ of item $i$ is inferred as follows:
\begin{equation}
\label{eq:enh}
    r^*_i =  \frac{1}{L+1}\sum_{l=0}^{L} r_i^{(l)},~~ \hat{r}_i = \frac{r^*_i}{||r_i^*||_2^2}
\end{equation}
where $L$ is the number of propagation layers in the GCN, and the final representation of item $i$ is obtained from $r_i^*$ followed by a second order Euclidean normalization.
Inspired by \cite{le2019correlation}, we calculate the correlation score $s_{i,j}$ of an item pair $(i, j), i, j \in \mathcal{V}$ as:
\begin{equation}
\label{eq:csim}
    s_{i,j} = \hat{r}_i^\intercal \cdot \hat{r}_j
\end{equation}
Given an item $v_i$, the item cluster $\mathcal{C}_i$  of $k$ nearest neighbors is determined using a distance function for each pair of items ($v_i$, $v_j$) as: 
\begin{equation}
\label{clus_d}
    d_{i,j} = \frac{1}{exp(s_{i,j})},
\end{equation}
\subsection{Pseudo Bundle Generation}
In order to facilitate the bundle recommendation task, we aim to construct a pseudo bundle, which expresses the preferential intention of a given user $u$. Inspired by \cite{yang2024generate}, it might be inferred via behavioral history $\mathcal{H}_u$, i.e., interacted items $\mathcal{V}_{\mathcal{H}_u} \in \mathcal{V}$. Using the clusters of those items built from the previous section, we suppose that they may create meaningful instructions to better construct the pseudo bundle. Given $\mathcal{B}_u \in \mathcal{B}$ as the bundle set associated with user $u$, $\mathcal{V}_u, \mathcal{V}_{\mathcal{B}_u} \in \mathcal{V}$ are the sets of adopted items of user $u$ extracted from $\mathcal{Z}$, and $\mathcal{B}_u$ respectively, we have: 
\begin{equation}
    \mathcal{V}_{\mathcal{H}_u} := \mathcal{V}_u \cup \mathcal{V}_{\mathcal{B}u}
\end{equation}


\subsubsection{\textbf{Instruction Construction.}} 
For each training iteration of a user $u$,  we randomly select an item $v_i$ from his historical interactions $\mathcal{V}_{\mathcal{H}_u}$. Using item cluster $\mathcal{C}_i$ of $k$ nearest neighbors of $v_i$ determined from the correlation-based item clustering module, an instructive bundle $b_{inst}$ is constructed via: 
\begin{equation}
  b_{inst} = \{\Bar{v}_{1}, \Bar{v}_{2}, \Bar{v
}_{3}, \dots, \Bar{v}_{k}\}
\end{equation} where it is noted that $1 <  k < n$, $n = |\mathcal{V}_{\mathcal{H}_u}|$, $k$ is randomly selected during training to add random noises for improving model robustness. Dependending on the number of historical interactions $n$, there may have a set of instructive bunbles $\mathcal{B}_{u, inst} = \{b_{inst}\}$ generated from multiple training iterations for each user $u$.  


\subsubsection{\textbf{Instruction-Driven Pseudo Bundle Generation.}} In order to form a pseudo bundle for relevant bundle retrieval, the instructive set $\mathcal{B}_{u, inst}$ is fed into a sequence-to-sequence architecture, e.g., Transformer, to aggregate potential items that highly correlate and align with user preferences. Motivated by \cite{yang2024generate}, we employ the reconstruction distribution process to reconstruct the temporal distribution of potential item probabilities. 
Using the collaborative signals from the user's previous interactions $ \mathcal{V}_{\mathcal{H}_u}$ and a given $b_{inst}$, the probability of pseudo bundle $b_{gen}$ after the $T-$th step generation is derived as follows:
\begin{align}
    p_\theta(\Tilde{v}^{(1:T)}) &:= \prod_{t=1}^T p_\theta(\Tilde{v}^{(t)} | \mathcal{V}_{\mathcal{H}_u}, \Bar{v}_{0:t-1}) \\
    p_\theta(\Tilde{v}^{(t)} | \mathcal{V}_{\mathcal{H}_u}, \Bar{v}_{0:t-1}) &:= \mathcal{N}( \Phi_\theta(\Psi_\theta(\mathcal{V}_{\mathcal{H}_u}),\Bar{v}_{0:t-1}), \beta \mathbf{I}), \notag
\end{align}
where $\Tilde{v}^{(t)}$ is the $t-$th candidate item, and $\Tilde{v}^{(1:T)}$ is the candidate item set after $T$ generation steps for $b_{gen}$. Likewise, $\Bar{v}_{0:t-1}$ is the item set consisted of a start-of-bundle pseudo item $[sob]$ at the first index and the first $t-1$ items of $b_{ins}$. $\mathcal{N}$ denotes the Gaussian distribution, $\mathbf{I} \in \mathbb{R}^{|\mathcal{V}|}$ is the identity tensor, $\Psi_\theta(\cdot)$ and $\Phi_\theta(\cdot)$ respectively are Transformer encoder and decoder \cite{vaswani2017attention}, and $\beta$ is a hyper-parameter control the variance of the probability distribution.
The final pseudo bundle $b_{gen} = \{\Tilde{v}_1, \Tilde{v}_2, \ldots, \Tilde{v}_t\}$ is the $t$ candidate items extracted from the $\Tilde{v}^{(1:T)}$ set, where $v_{t+1}$ is the end-of-bundle pseudo item $[eob]$. 

\textit{During inference process}, we do not utilize the instructive bundle set to neglect biases in constructing the pseudo bundle. It is merely generated based on the historical interaction of users $\mathcal{V}_{\mathcal{H}_u}$ and previously-selected candidate items via the following procedure:
\begin{align}
    p_\theta(\Tilde{v}^{(1:T)}) &:= \prod_{t=1}^T p_\theta(v^{(t)} | \mathcal{V}_{\mathcal{H}_u}, \Tilde{v}^{(0:t-1)}), \\
    p_\theta(\Tilde{v}^{(t)}|v_{1:n}, \Tilde{v}^{(0:t-1)}) &:= \mathcal{N} (\Phi_\theta(\Psi_\theta(\mathcal{V}_{\mathcal{H}_u}), \Tilde{v}^{(0:t-1)}), \beta_I \mathbf{I}), \notag
\end{align} where the final pseudo bundle $b_{gen}$ is built in the similar way as the training phase. 
\subsection{Retrieval \& Ranking}
With the objective of recommending existing bundles, we employ a retrieval and ranking workflow. The main idea is to discover for the top$-K$ bundles in $\mathcal{B}$ that are most similar to the pseudo bundle $b_{gen}$. It raises a need to calculate similarity score between the pseudo bundle $b_{gen}$ and each bundle $b \in \mathcal{B}$ of a given user $u$. Generally, there are two typical similar metrics including Jaccard similarity $y_{u, b}^J$ and Cosine similarity $y_{u, b}^C$. The former favors the exact matching among bundles while the latter emphasizes the relative one using latent preferential features. 

For Jaccard similarity, let us denote $c_{g}, c_{b} \in \{0,1\}^{|\mathcal{V}|}$ are binary vectors that represent the occurrence of items within the pseudo bundle $b_{gen}$ and a bundle $b \in \mathcal{B}$, we have:
\begin{equation}
    y^J_{u,b} = \frac{{c}_{g}^\intercal \cdot {c_b}}{c_{g}^\intercal \cdot \mathbf{I} + c_b^\intercal \cdot \mathbf{I} - c_{g}^\intercal \cdot c_b}, \\
\end{equation} 

Equally important, the cosine similary $y_{u, b}^C$ is computed as the following procedure: 
\begin{align}
    \hat{b} &= \frac{1}{c_b^\intercal \cdot \mathbf{I}}  c_b^\intercal \cdot \mathcal{\hat{R}} \\ 
    \hat{b}_{gen} &= \frac{1}{c_{g}^\intercal \cdot \mathbf{I}} c_{g}^\intercal \cdot \mathcal{\hat{R}} \\
    y_{u,b}^C &= \frac{\hat{b}^\intercal \cdot \hat{b}_{gen}}{||\hat{b}||^2_2 \cdot ||\hat{b}_{gen}||^2_2}
\end{align} where $\mathcal{\hat{R}} \in \mathbb{R}^{|\mathcal{V}| \times d}$ is the latent item representation obtained from the correlation-based item clustering module via Eq~(\ref{eq:enh}). Finally, we combine the two metrics to leverage all relevant bundles considering both two matching strategies:
\begin{equation}
    y_{u,b} = \alpha.y_{u, b}^J + (1-\alpha)y_{u, b}^C
\end{equation} where $\alpha \in [0, 1] $ is a trade-off hyperparameter to control the balance between two terms.The top-$K$ candidate bundles with the highest similarity score $y_{u,b}$ are recommended to the user $u$.

\subsection{Optimization}

Our model \mName{} is trained with triplet losses namely clustering loss $\mathcal{L}_{C}$, the pseudo bundle generation loss $\mathcal{L}_{G}$ and the recommendation loss $\mathcal{L}_{R}$. Specifically, the clustering loss is to maximize the correlation score between potential items within bundle instructions. It is calculated as:
\begin{equation}
    \mathcal{L}_{C} = \sum_{(v_i, v_j, v_{j'}) \in P} -ln~\sigma(ln(\frac{1}{d_{i,j}}) - ln(\frac{1}{d_{i,j'}})),
\end{equation}
where $\sigma(\cdot)$ denotes the Sigmoid function. 
$d_{i,j}$ represents the distance between item pair ($v_i, v_j$) in the latent space as Eq~(\ref{clus_d}), and $P = \{(v_i,v_j,v_{j'}) | v_i,v_j, v_{j'} \in \mathcal{V}, C_{i,j}=1, v_i \neq v_j, C_{i,j'}=0, v_i \neq v_{j'} \}$ 




To distill the knowledge from the instruction-driven bundles to the pseudo bundle, we apply the cross-entropy loss over $T$ timesteps as:
\begin{equation}
    \mathcal{L}_{G} = -\frac{1}{T}\sum_{t = 1}^T \ln(p_\theta(r^t~|~v_{1:n}, r_{0:t-1})^\intercal) \cdot b_{inst}^{(t)},
\end{equation}
where $b_{inst}^{(t)}$ is the target distribution at $t$ given by $b_{inst}$. 

Inspired by \cite{rendle2012bpr}, the bundle recommendation loss is computed as:
\begin{equation}
    \mathcal{L}_R = \sum_{(u,b, b') \in Q} 
    -ln\sigma(y_{u,b} - y_{u, b'}),
\end{equation}
where $Q = \{(u,b,b')|u \in \mathcal{U}; b,b'\in \mathcal{B}; Z_{u,b}=1, Z_{u,b'}=0\} $. Finally, the combined loss function of \mName{} is achieved as follows: 
\begin{equation}
    \mathcal{L} =  \mathcal{L}_{G} + \mathcal{L}_{C} + \mathcal{L}_R + \lambda||\theta||^2_2,
\end{equation}
where $||\theta||^2_2$ denotes the $L2$ regularization, and $\lambda$ indicates a hyperparameter to control.

\begin{table*}[t]
\renewcommand{\arraystretch}{0.8}
\centering
\scalebox{0.76}{
\begin{tabular}{c|c|cc|cccccccc|c|c}
\toprule
\multicolumn{1}{c|}{\textbf{Dataset}} & \multicolumn{1}{l|}{\textbf{Metric}} & \multicolumn{1}{l}{BPRMF} & \multicolumn{1}{l|}{DAM} & \multicolumn{1}{l}{LightGCN} & \multicolumn{1}{l}{BGCN} & \multicolumn{1}{l}{MIDGN} & \multicolumn{1}{l}{CrossCBR} & \multicolumn{1}{l}{MultiCBR} & \multicolumn{1}{l}{CoHeat} & \multicolumn{1}{l}{BundleGT} & \multicolumn{1}{l|}{BunCa} & \multicolumn{1}{l|}{\textbf{BRIDGE}} & \multicolumn{1}{l}{Imp($\%$)} \\ \midrule
\multirow{4}{*}{\textbf{Clothing}} 
& \textbf{R@1} & 0.0126 & 0.0174 & 0.0213 & 0.0616 & 0.1057 & 0.6806 & 0.5775 & \underline{0.7162}& 0.3496 & 0.7043 & $\textbf{0.8511}^{\dagger}$ & 18.8\%\\
& \textbf{R@2} & 0.0282 & 0.0354 & 0.0426 & 0.0937 & 0.1635 & 0.8334 & 0.6995 & {0.8461} & 0.3942 & \underline{0.8594}& $\textbf{0.9871}^{\dagger}$ & 16.6\%\\
& \textbf{N@1} & 0.0142 & 0.0216 & 0.0328 & 0.0669 & 0.1324 & 0.7356 & 0.6211 & \underline{0.7741}& 0.3667 & 0.7560 & $\textbf{0.9248}^{\dagger}$ & 19.3\%\\
& \textbf{N@2} & 0.0266 & 0.0305 & 0.0574 & 0.0836 & 0.1781 & 0.7987 & 0.6704 & {0.8186} & 0.3823 & \underline{0.8212}& $\textbf{0.9653}^{\dagger}$ & 17.9\%\\ \midrule
\multirow{4}{*}{\textbf{Electronic}} 
& \textbf{R@1} & 0.0214 & 0.0135 & 0.0337 & 0.0533 & 0.0781 & 0.5528 & 0.4632 & \underline{0.7058}& 0.2779 & 0.6566 & $\textbf{0.8451}^{\dagger}$ & 19.8\%\\
& \textbf{R@2} & 0.0275 & 0.0567 & 0.0645 & 0.0672 & 0.1229 & 0.7348 & 0.5917 & \underline{0.8422}& 0.3555 & 0.8044 & $\textbf{0.9673}^{\dagger}$ & 14.8\%\\
& \textbf{N@1} & 0.0217 & 0.0135 & 0.0345 & 0.0557 & 0.1036 & 0.5907 & 0.4968 & \underline{0.7634}& 0.2981 & 0.7582 & $\textbf{0.9055}^{\dagger}$ & 18.6\%\\
& \textbf{N@2} & 0.0247 & 0.0324 & 0.0535 & 0.0646 & 0.1621 & 0.6842 & 0.5554 & \underline{0.8132}& 0.3351 & 0.7693 & $\textbf{0.9464}^{\dagger}$ & 15.7\%\\ \midrule
\multirow{4}{*}{\textbf{Food}} 
& \textbf{R@1} & 0.0105 & 0.0124 & 0.0193 & 0.0822 & 0.0961 & 0.5665 & 0.4986 & \underline{0.7314}& 0.3177 & 0.6401 & $\textbf{0.8335}^{\dagger}$ & 13.9\% \\
& \textbf{R@2} & 0.0242 & 0.0210 & 0.0372 & 0.1052 & 0.1986 & 0.7184 & 0.6305 & \underline{0.8663}& 0.4314 & 0.7655 & $\textbf{0.9465}^{\dagger}$ & 9.2\% \\
& \textbf{N@1} & 0.0133 & 0.0141 & 0.0215 & 0.0887 & 0.1238 & 0.6216 & 0.5399 & \underline{0.7996}& 0.3453 & 0.7019 & $\textbf{0.9041}^{\dagger}$ & 13.0\%\\
& \textbf{N@2} & 0.0253 & 0.0154 & 0.0307 & 0.0988 & 0.3034 & 0.6821 & 0.5966 & \underline{0.8423}& 0.4028 & 0.7392 & $\textbf{0.9466}^{\dagger}$ & 12.3\%\\ \midrule
\multirow{4}{*}{\textbf{Steam}} 
& \textbf{R@1} & 0.0033 & 0.0016 & 0.0017 & 0.0014 & 0.0022 & 0.0410 & 0.0664 & \underline{0.2351}& 0.0014 & 0.0489 & $\textbf{0.3472}^{\dagger}$ & 43.4\%\\
& \textbf{R@2} & 0.0046 & 0.0025 & 0.0013 & 0.0218 & 0.0267 & 0.0995 & 0.1139 & \underline{0.3435}& 0.0124 & 0.1253 & $\textbf{0.4298}^{\dagger}$ & 20.6\%\\
& \textbf{N@1} & 0.0032 & 0.0023 & 0.0017 & 0.0044 & 0.0058 & 0.0859 & 0.1112 & \underline{0.2913}& 0.0013 & 0.0897 & $\textbf{0.4504}^{\dagger}$ & 54.6\%\\
& \textbf{N@2} & 0.0032 & 0.0043 & 0.0012 & 0.0195 & 0.0212 & 0.1124 & 0.1344 & \underline{0.3336}& 0.0075 & 0.1350 & $\textbf{0.4532}^{\dagger}$ & 32.7\%\\ \bottomrule
\end{tabular}}
\caption{Overall performances on five benchmark datasets.
The best results are in \textbf{bold}, and the second best results are {underlined}. The symbol $\dagger$ indicates statistically significant improvements over the second-best models with ($p < 0.01$)}
\label{compare_table}
\end{table*}

\section{Experiments}



\subsection{Experimental Setup}
\subsubsection{\textbf{Datasets.}} We run extensive experiments on five datasets in multiple domains namely Clothing, Electronics, Food and Steam. The data statistics are illustrated in Table \ref{data_stat_table}.

\begin{itemize}
    \item Clothing, Electronic, Food \cite{sun2022revisiting} are constructed from the Amazon dataset with high quality bundles of products using crowd-sourcing resources.
    \item Steam\footnote{\url{http://cseweb.ucsd.edu/∼jmcauley/}} \cite{pathak2017generating} includes bundles of games purchased together on the Australian game platform.
    \item iFashion \cite{chen2019pog} is a fashion outfit recommendation dataset, each outfit is treated as a bundle.
\end{itemize}

\begin{table}[t]
\centering
\scalebox{0.82}{
\begin{tabular}{l|c|c|c|c}
\toprule
\textbf{Dataset}      &\textbf{Clothing} & \textbf{Electronic} & \textbf{Food} & \textbf{Steam} 
\\ \midrule
\#User       &$965$& $888$& $879$&$29,634$
\\
\#Item       &$4,487$& $3,499$& $3,767$&$2,819$
\\
\#Bundle     &$1,910$& $1,750$& $1,784$&$615$
\\ \midrule
U-I Density  & $0.15\%$& $0.20\%$& $0.19\%$&$1.08\%$
\\
U-B Density  & $0.10\%$& $0.11\%$& $0.11\%$&$0.48\%$
\\
Avg \#I/B    & $3.31$& $3.52$& $3.58$&$5.73$
\\ 
Avg \#B/I    & $1.40$& $1.76$& $1.69$&$1.25$
\\
Avg $|\mathcal{V}_{\mathcal{H}_u}|$ &10.72&11.25&11.80&37.60
\\ \bottomrule
\end{tabular}}
\caption{Statistics of benchmark datasets.}
\label{data_stat_table}
\end{table}

\subsubsection{\textbf{Comparative Models.}}
To demonstrate the effectiveness, we compare BRIDGE with several state-of-the-art models for bundle recommendation: 
\begin{itemize}
    \item Factorization Models: BPRMF \cite{rendle2012bpr} and DAM \cite{chen2019matching};
    \item Graph-based Models: LightGCN \cite{he2020lightgcn}, 
    BGCN \cite{chang2020bundle}, 
    MIDGN \cite{zhao2022multi}, 
    CrossCBR \cite{ma2022crosscbr}, 
    BundleGT \cite{BundleGT2023}, 
    MultiCBR \cite{ma2024multicbr},
    CoHeat \cite{jeon2024cold}
    and BunCa \cite{nguyen2024bundle}.

\end{itemize}

\subsubsection{\textbf{Evaluation Strategy \& Metrics.}}
To ensure a fair comparison, we divide the data into training, validation, and test sets using a $7$:$1$:$2$ ratio, consistent with the comparative models. For performance evaluation, we utilize two typical metrics for the Top-$K$ recommendation task: Recall ($R@K$) and Normalized Discounted Cumulative Gain ($N@K$). Specifically, $R@K$ reflects the ratio of true recommended bundles to all ground truth bundles, while $N@K$ is higher when the true recommended bundles appear at the top of the retrieval ranking list. For both metrics, we report the average performance of all models on the test bundles using $5$ runs with different random initializations. Significant differences are validated using a two-tailed paired-sample Student's t-test at a $0.01$ significance level.

\subsubsection{\textbf{Implementation Details.}}
All comparative models are reproduced using their official source code
while \mName{} is implemented using Flax\footnote{\url{https://flax.readthedocs.io/}}. Our model is optimized with Adam optimizer \cite{kingma2014adam} with the learning rate tuned in the set of $\{1e-3, 1e-4, 1e-5\}$. The number of encoder/decoder blocks, attention heads, and graph encoder layers are empirically selected in the range of values $\{1, 2, 3, 4\}$. For top-$K$ recommendation, we select $K \in \{1, 2\}$ for all datasets to favour the highly-ranked bundles. All experiments are run on a single NVIDIA P100 GPU.

\begin{table*}[t]
\centering
\scalebox{0.75}{\begin{tabular}{c|cccc|cccc}
\toprule
Dataset  & \multicolumn{4}{c|}{Steam}    & \multicolumn{4}{c}{Electronic}      \\ \midrule
Metric   & R@1   & R@2   & N@1   & N@2   & R@1  & R@2  & N@1  & N@2  \\ \midrule
BRIDGE      & $\textbf{0.342}$ & $\textbf{0.429}$ & $\textbf{0.450}$ & $\textbf{0.453}$  & $\textbf{0.845}$ & $\textbf{0.967}$ & $\textbf{0.905}$ & $\textbf{0.946}$ \\
w/o Inst & $0.275_{\downarrow19.5\%}$ & $0.326_{\downarrow24.0\%}$ & $0.365_{\downarrow18.8\%}$ & $0.357_{\downarrow21.1\%}$ 
& $0.702_{\downarrow16.9\%}$ & $0.829_{\downarrow14.2\%}$ & $0.762_{\downarrow15.8\%}$ & $0.804_{\downarrow15.0\%}$  \\

w/o Gen & $0.071_{\downarrow 79.2\%}$ & $0.128_{\downarrow70.1\%}$ & $0.107_{\downarrow76.2\%}$ & $0.132_{\downarrow70.8\%}$ 
& $0.394_{\downarrow53.3\%}$ & $0.439_{\downarrow54.6\%}$ & $0.428_{\downarrow52.7\%}$ & $0.434_{\downarrow54.1\%}$ \\ 
\bottomrule
\end{tabular}}
\caption{Impact of key components on the performance of \mName{}.}
\label{ablation_table}
\end{table*}

\subsection{Comparisons with Comparative Models}


Table~\ref{compare_table} presents the performance comparison between \mName{} and other comparative models. Conventional recommendation techniques, such as BPRMF and LightGCN, are less effective for bundle recommendation because they focus on optimizing representations of users and bundles but fail to capture item-level information adequately. In contrast, state-of-the-art methods like CrossCBR, MultiCBR, and CoHeat perform well across all five datasets by modeling user preferences at the item level and minimizing inconsistencies between different levels of preferences through multi-view learning. Our model, \mName{}, consistently outperforms all comparative methods across both evaluation metrics on the five benchmark datasets, demonstrating its effectiveness for bundle recommendation.

For Steam, \mName{} shows significant improvements from $20.6\%$ to $54.6\%$ compared to the second-best approach. 
It could be explained by the statistics of Steam dataset, in which the average number of items in a bundle is $5.73$, and each item appears in an average of $1.25$ bundles. This implies that any two randomly selected bundles are likely to have relatively low similarity, as they will contain different combinations of the available items. 
If the pseudo bundle has just one item matching the unseen truth bundles, it increases the probability to retrieve the truth bundle. 


\subsection{Model Component Contribution}

We conduct various ablation studies to investigate the importance of \mName{} main components. We present the analysis for the Steam and Electronic datasets, with the remaining results provided in the appendix.

The ablation results are shown in Table~\ref{ablation_table}. Specifically, for the w/o Inst scenario, instructions are removed, meaning the generation module is trained without any guidance from the instructions. In the w/o Gen scenario, instead of generating pseudo ideal bundles, we aggregate all user instructions as user preferences.
\textbf{Without guidance from instructions} (w/o Inst), there is a significant drop in performance on both the Steam and Electronic datasets: $R@1$ decreases by $19.5\%$, $R@2$ by $24\%$, $N@1$ by $18.8\%$, and $N@2$ by $21.1\%$ on Steam. On Electronic, the reductions range from $14\%$ to $17\%$.
This shows the importance of training the bundle generation within the correlation guidance signals. Without it, not only does it affect the quality of the generated bundles, but it also drags down the similarity between pseudo bundle and predefined ones due to introducing more noisy items to the `ideal' bundles.
The effectiveness of the generation component is also highlighted. \textbf{Without pseudo bundle generation } (w/o Gen), a substantial drop in performance is consistently observed, with reductions exceeding $70\%$ on Steam and over $50\%$ on Electronic.

\begin{figure}[t]
\centering
\includegraphics[width=0.42\textwidth]{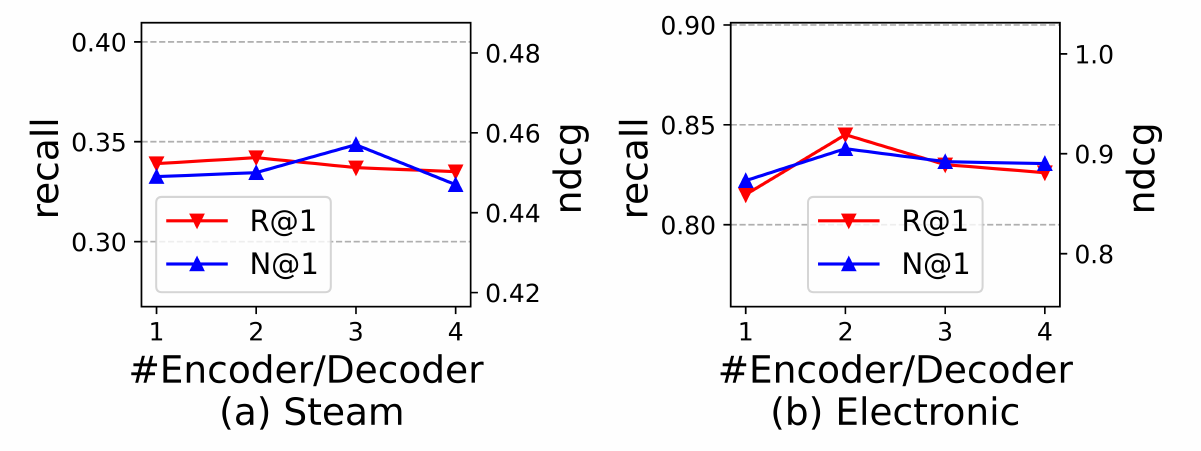}
\caption{Impact of $\#$Encoder/Decoder block.}
\label{block_fig}
\end{figure}

\begin{figure}[t]
\centering
\includegraphics[width=0.42\textwidth]{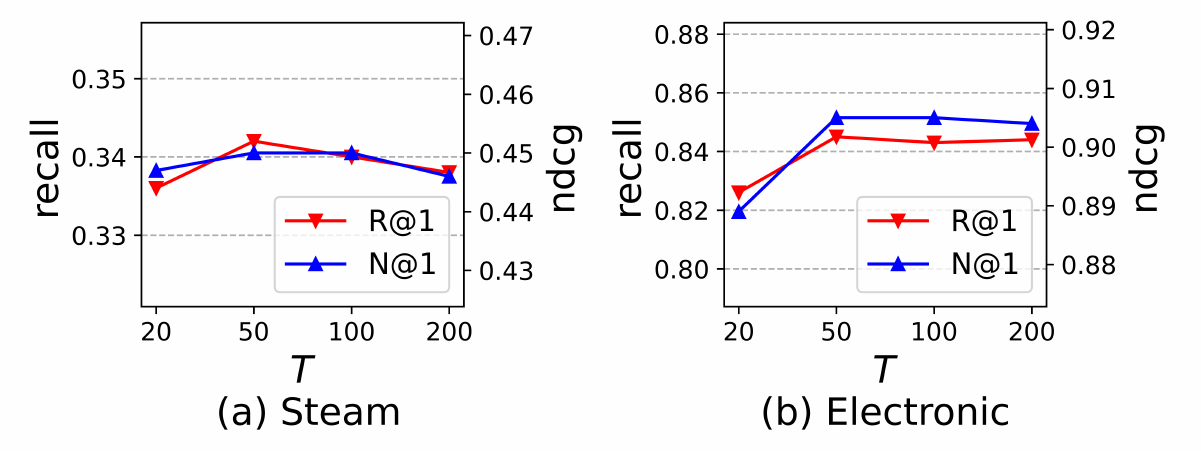}
\caption{Impact of max-context length $T$.}
\vspace{-10px}
\label{context_len_fig}
\end{figure}

We also conduct experiments to further analyze the impact of the number of encoder-decoder layers and the maximum context length $T$ on \textit{the Pseudo Bundle Generation module}.
As shown in Figure~\ref{block_fig}, the \textbf{number of encoder-decoder layers} noticeably impacts model performance. Increasing the number of encoder/decoder blocks from $1$ to $2$ leads to a clear improvement on Electronic, while performance on Steam remains stable across different numbers of layers.
The \textbf{maximum context length $T$} has varying effects on results across different datasets. Figure~\ref{context_len_fig}a shows that \mName{} maintains stable performance on Steam across different values of $T$, while $T \geq 50$ performs better than $T=20$ on Electronic (see Figure~\ref{context_len_fig}b). This is because a longer context length captures more aspects of user historical interactions. 
On Steam, where the average number of past interactions is $37.60$, $T=20$ is insufficient for capturing user collaborative signals. On Electronic, with an average of $11.25$ interactions and $8.8\%$ of users having more than $20$ past interactions, $T=20$ results in performance degradation. With $T \geq 50$, \mName{} effectively captures all user preferences, leading to more stable performance.

\begin{figure}[t]
\centering
\includegraphics[width=0.35\textwidth]{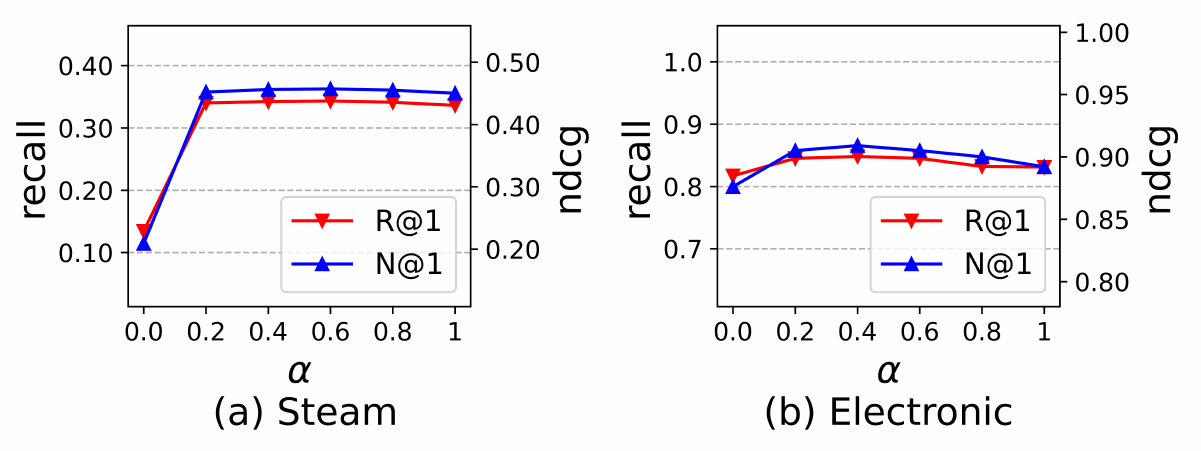}
\vspace{-5px}
\caption{Impact of trade-off coefficient $\alpha$.}
\label{alpha_fig}
\end{figure}

In the \textit{Retrieval \& Ranking module}, we investigate the impact of the \textbf{trade-off coefficient} \textbf{$\alpha$} between two similarities on the model performance as described in Figure~\ref{alpha_fig}. 
Without the combination of two similarity metrics, i.e., $\alpha = 0$ or 1, the \mName{}'s performance drops considerably. On Steam dataset, $\alpha=0$ makes a sharp drop in performance, while with $\alpha=1$, $R@1$ decreases from $0.342$ to $0.336$. 
The results on Electronic also show a noticeable decrease in performance for both $\alpha=0$, and $\alpha=1$, highlighting the importance of our combined similarity metric for retrieving bundles.

Our model generates a single pseudo `ideal' bundle for each user but struggles to recommend multiple bundles with varied interests using Jaccard similarity. The model is also experimented with using cosine similarity when retrieving the recommendation list. This allows the generated bundles to cover a broader range of aspects that may be relevant to the user. But this comes at the potential cost of losing the `ideal' nature of the bundle, as the focus shifts more towards diversity rather than optimization for a single aspect. The trade-off between accuracy and aspect diversity is shown in Table~\ref{tradeoff_table}. 
With small values of top-K, BRIDGE performs better with Jaccard, whereas with larger values of top-K, the accuracy when retrieving with Jaccard similarity becomes lower than Cosine similarity.
To overcome the trade-off issue, we combine two retrieval strategies to obtain the best result over all the retrieval recommendation size.


\begin{table}[t]
\centering
\scalebox{0.67}{
\begin{tabular}{c|c|c|c|c|c|c}
\toprule
R@K     & $K=1$   & $K=2$   & $K=5$   & $K=10$  & $K=20$  & $K=50$  \\ \midrule
\multicolumn{1}{r|}{Jaccard Only ($\alpha$ = 1)} & \underline{0.336} & \underline{0.415} & \underline{0.741} & \underline{0.752} & \underline{0.752} & 0.780 \\
\multicolumn{1}{r|}{Cosine Only ($\alpha$ = 0)} & 0.134 & 0.169 & 0.227 & 0.360 & 0.646 & \underline{0.961} \\ 
\midrule
\mName{} &\textbf{0.342}&\textbf{0.429}&\textbf{0.748}&\textbf{0.754}&\textbf{0.800}&\textbf{0.979}\\
\bottomrule
\end{tabular}
}
\caption{Performance of BRIDGE on Steam over size of retrieval recommendation list.}
\label{tradeoff_table}
\end{table}

\subsection{Qualitative Analysis}
The quality of a generated pseudo ideal bundle is hard to evaluate. \citet{pathak2017generating, han2017learning} focused on evaluating the compatibility score between items within the generated bundles as a way to measure their quality.
\citet{chen2019pog} experimented the generated bundles on an online platform which is hard to follow.
To address this problem, we evaluate our generated bundles through a downstream bundle recommendation task. 
A generated bundle is denoted as a meaningful bundle if it highly similar to an unseen bundle of a user.

\begin{figure}[t]
\centering
\includegraphics[width=0.42\textwidth]{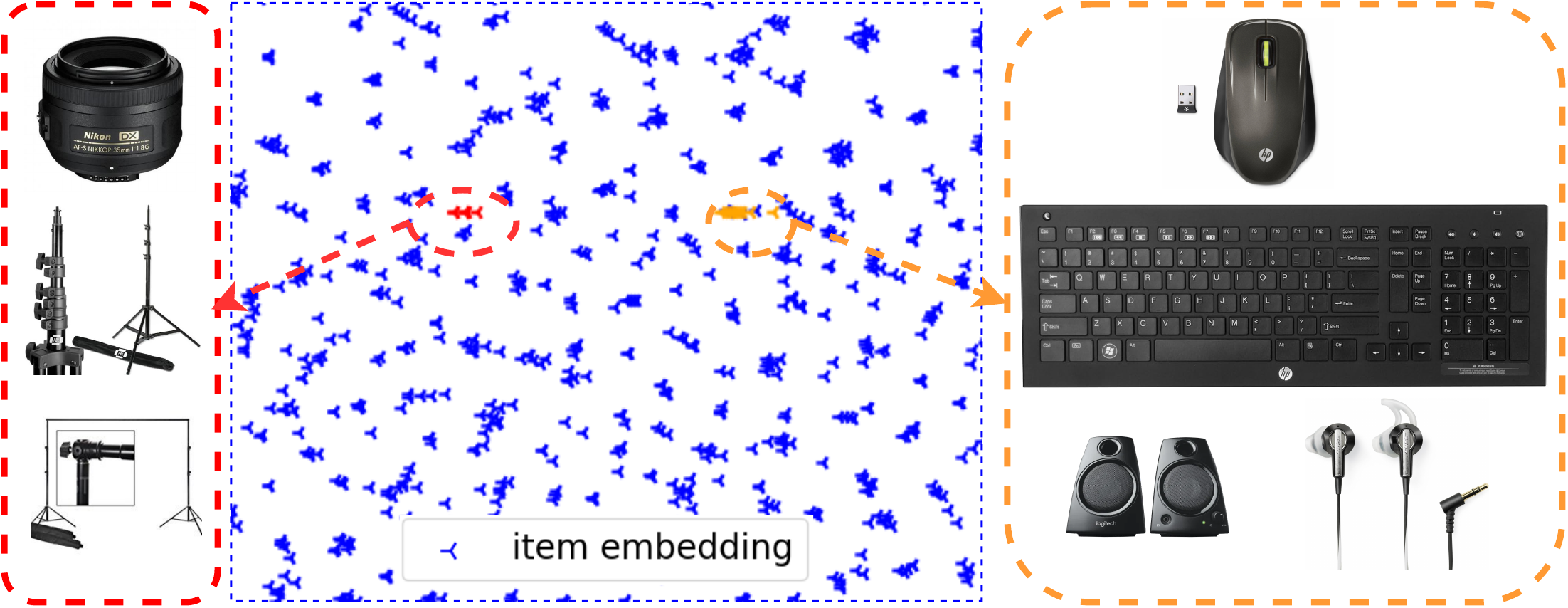}
\caption{T-SNE visualization of correlation-based instructions on Electronic.}
\label{tsne_cls_ele_fig}
\vspace{-8px}
\end{figure}

\begin{figure}[t]
\centering
\includegraphics[width=0.40\textwidth]{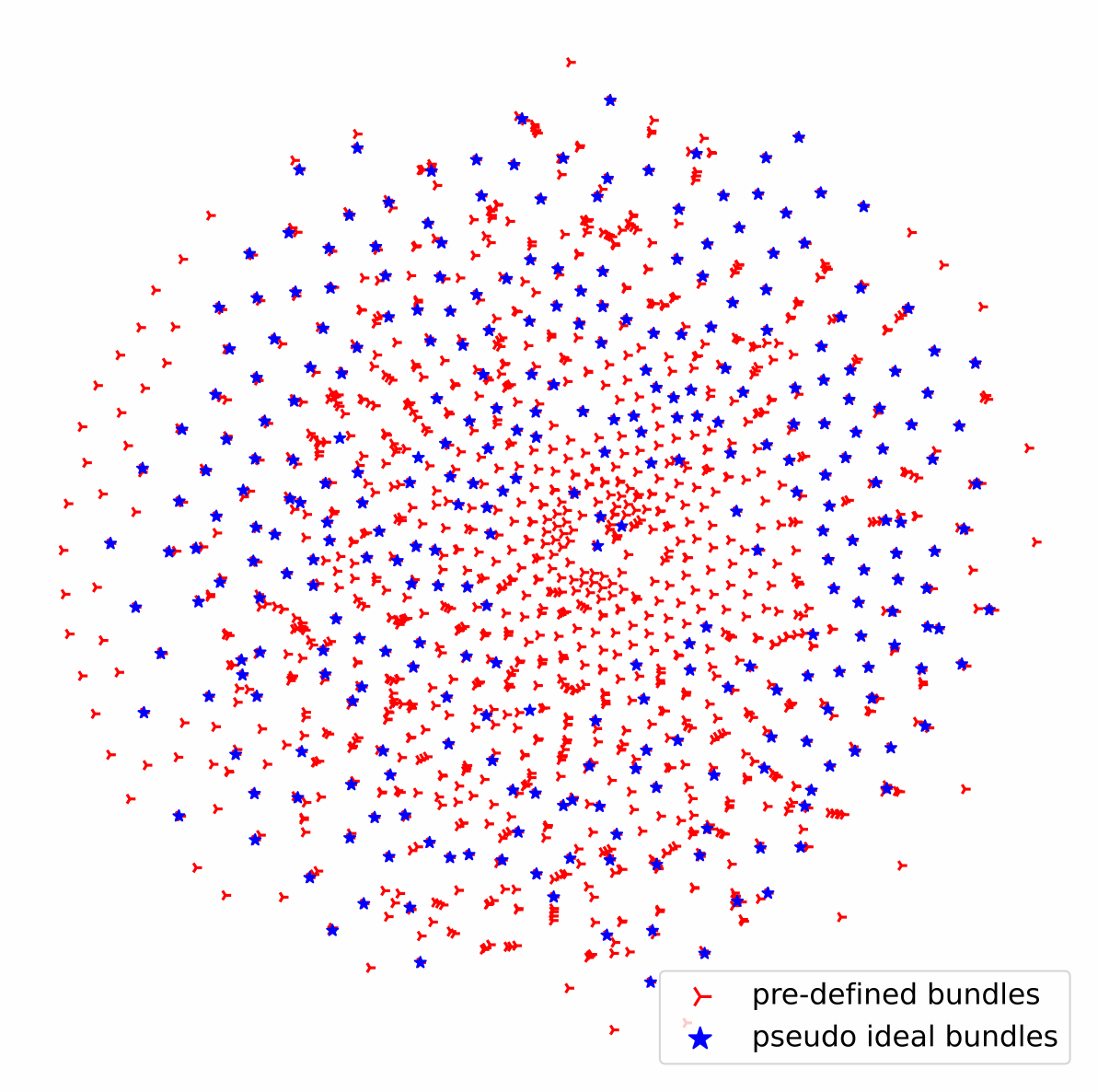}
\caption{T-SNE visualization of pseudo ideal bundles and ground-truth ones on Electronic.}
\label{gen_ideal_bun_fig}
\end{figure}

\textbf{Correlation-based Item Cluster}. Before demonstrating that the generated bundles are meaningful, we show some examples of the instructions given to the generator are consistent wherein the items have same underlying representation. 
Figure~\ref{tsne_cls_ele_fig} shows the latent representations for items where items with high correlation are close to each other, some of the meaningful clusters are shown: the ``\textit{Desktop}'' cluster contains ``\textit{Logitech Speakers}'', ``\textit{Audio Headphones}'', ``\textit{Wireless Keyboard}'', and ``\textit{Wireless Mouse}''.
While the ``\textit{Camera}'' cluster is a combination of ``\textit{Lens for Digital Cameras}'', ``\textit{Light Stand}'' and ``\textit{Backdrop Background}''.

\textbf{Pseudo `Ideal' Bundle}. For user in the testing set, the output bundles generated by our model are highly similar with the ground truth bundles at item level. 
This leads to very significant improvement over all baseline methods,
on Steam the generated bundle tends to be a jointly subset of ground truth bundles that make BRIDGE can retrieve multiple bundles with different aspects which match user varied interest. 
The product bundles generated by the BRIDGE model have been shown to exhibit meaningful and useful combinations of items, as illustrated in Figure~ \ref{gen_ideal_bun_fig}.
The occurrence of items in a generated bundles is similar compare to the bundles in retrieval recommendation list of users. 
Additionally, the distribution of the generated bundles is similar to the distribution of the ground truth bundles, which demonstrates that the generated bundles cover all the aspects of the pre-constructed bundles in a diverse manner. 



\section{Conclusion}
In this paper, we introduce \mName{}, a novel framework for bundle recommendation inspired by distant supervision strategies and the generative paradigm. The framework integrates correlation-based item clustering modules that enables the generation of auxiliary information without relying on external data. By combining this with collaborative signals from user historical interactions, \mName{} can produce pseudo 'ideal' bundles, which allows to explore a broader range of potential bundles beyond those predefined ones. This approach effectively narrows the gap between user expectations and available bundles, leading to improved recommendation performance. Extensive experiments demonstrate the effectiveness of \mName{} over comparative models for bundle recommendation on five public datasets.

\nobibliography*




\bibliography{aaai25}
\appendix

\end{document}